\documentclass{amsart}
\usepackage{graphicx}
\def\euro{\mbox{\raisebox{.25ex}{{\it =}}\hspace{-.5em}{\sf C}}}
\begin{document}
\author{F. d'Acapito}
\address[F. d'Acapito]{CNR-IOM-OGG, 71 Avenue des Martyrs, Grenoble France}
\author{A. Trapananti}
\address[A. Trapananti]{CNR-IOM Perugia c/o Dipartimento di Fisica e Geologia- Univ. Perugia Via A. Pascoli, Perugia, Italy } 
\date{\today}
\begin{abstract}
This document presents the refurbishment project for GILDA, the Italian CRG beamline at the ESRF. After a short introduction to the beamline and its achievements in the last years, the base ideas and concepts driving the design are exposed whereas a more detailed technical description of the new instrument is given in a further section. An analysis of the costs and a schedule for the implementation of the program is also presented in the last part of the document. 
\end{abstract}
\title{Project of refurbishment of the GILDA beamline at the ESRF}
\maketitle
\tableofcontents
\clearpage
\section{Introduction}
The GILDA project was proposed in 1989 to provide the Italian community with an easy access to a III generation synchrotron radiation facility, ESRF. The main scientific goal was the study of structure of materials to be conducted via the X-ray Absorption and Diffraction techniques. The principal technical features were an high monochromatic flux, a small spot size and wide energy range. The beamline was installed in 1993 and the first user experiments were carried out in autumn 1994. Since then the experiments carried out by the users have produced about 550 publications in International reviewed journals with an average Impact Factor of about 3.0 \cite{gilda-report-2009}.\\
After many years of continuous operation a refurbishment of several beamline components is needed, in order to keep the beamline performance at the state-of-the-art, improve its reliability and to keep it at the leading edge among the other similar beamlines at the ESRF as well as in the world. It is worth noting that in the last years ESRF has undergone a robust refurbishment plan of its beamlines \cite{esrf-purple-book} As a matter of fact, during the last two decades relevant technological advances in the field of X-ray optics (mirror and monochromator) have been achieved and a modification of the beamline optics components will allow to improve the beam quality (size, stability and homogeneity) fully exploiting the extreme brilliance of the present ESRF X-ray beam and that of the future machine planned for the next years \cite{esrf-white-paper}. At present, the reliability of some components, in particular of the monochromator, is becoming a critical issue and such aspects have been also underlined by the last ESRF Review Committee (2009). In his report the committee appreciated the scientific and technical performances and achievements of GILDA and suggested to ESRF to endorse the project, but recommended a substantial refurbishment of the beamline.
\\
In the renewed version presented in this document, the limitations of the present layout (beam inhomogeneity, poor temporal stability) will be overcome and new experimental possibilities will be made available to users. This will make GILDA a unique tool for the advanced analysis of materials in a variety of field like Environmental Science, Materials science, Biophysics, Chemistry, Earth Science and Cultural Heritage.

\section{Conceptual design}
\label{sec:project}
It is worth to remind that GILDA is a unique instrument for the Italian and international community because it makes available to users an intense photon flux in a high energy range (namely, $> 20$ keV) that is only obtainable at the ESRF. The peculiar experimental aspects of the refurbished beamline, that will make GILDA a top-level instrument, are exposed below:
\begin{itemize}
\item use of grazing incidence/total reflection data collection in linear dichroic mode for surface analysis;
\item high sensitivity for the analysis of diluted samples;
\item high quality in term of low noise and linearity for XAS data in transmission mode.
\item  combined multitechnique \textit{in situ} data collection for extreme conditions or  \textit{in operando} experiments;
\end{itemize}
To achieve these targets a particular attention will be devoted to the quality of the beam spot in terms of homogeneity, spatial and energy stability, reduced size and divergence. For the scientific community it will be a fundamental tool to add value to the research carried out in home laboratories opening the perspective of high impact research.
To carry out the outlined research programs,  the beamline has to fulfill the following requirements:
\begin{itemize}
\item to provide a beam also at high energies (E $> 20$ KeV). This will permit the structural analysis (using XAS at the K absorption edge) of technologically crucial materials like catalysts (containing Mo, Pd), hydrogen reservoirs (Nb, Pd), protonic conductors (In, Ba, Ce, Zr), solar cells and transparent conductors (In,Sn), luminescent and advanced magnetic materials (Rare Earths).
\item To realize experiments in Grazing Incidence and linear dichroic mode. This will permit the structural analysis of thin films, interfaces or the interaction between adsorbed species and model surfaces, as needed for example in magnetic materials or environmental science.
\item To provide an intense and sub-mm beam in the whole energy range. This is needed for the analysis of diluted materials, for the studies under (moderate) extreme conditions of high temperature or pressure and for pump-probe experiments with electric field or light excitation. 
\item To provide a beam with reduced divergence, in order to carry out experiments based on X-ray reflectivity.
\end{itemize}
The refurbishment of the beamline can be realized in two phases. The objective of the first phase is the renewal of the X-ray optics and the adaptation of the infrastructure to the new beamline layout. The second phase will will be devoted to the installation of a new experimental setup with related measurement chambers and a new detector for X-ray fluorescence. The unique aspects that will have a major impact for the scientific community using this instrument will be:
\begin{itemize}
\item the possibility of realizing surface analysis with ReflEXAFS coupled to linear dichroism (i.e. with the beam polarization parallel or perpendicular to the surface). 
\item experiments on diluted samples also in extreme conditions and on the K absorption edges of heavy (namely 4d metals) elements.
\item experiments in pump-probe mode using the unique time structure of the beam (16 bunch, 4 bunch) available at ESRF. 
\item realization of coupled multi-technique in situ experiments.
\end{itemize}
\section{Technical design}
\subsection{Layout}
\begin{figure}
\begin{center}
\includegraphics[width=7cm, angle=90] {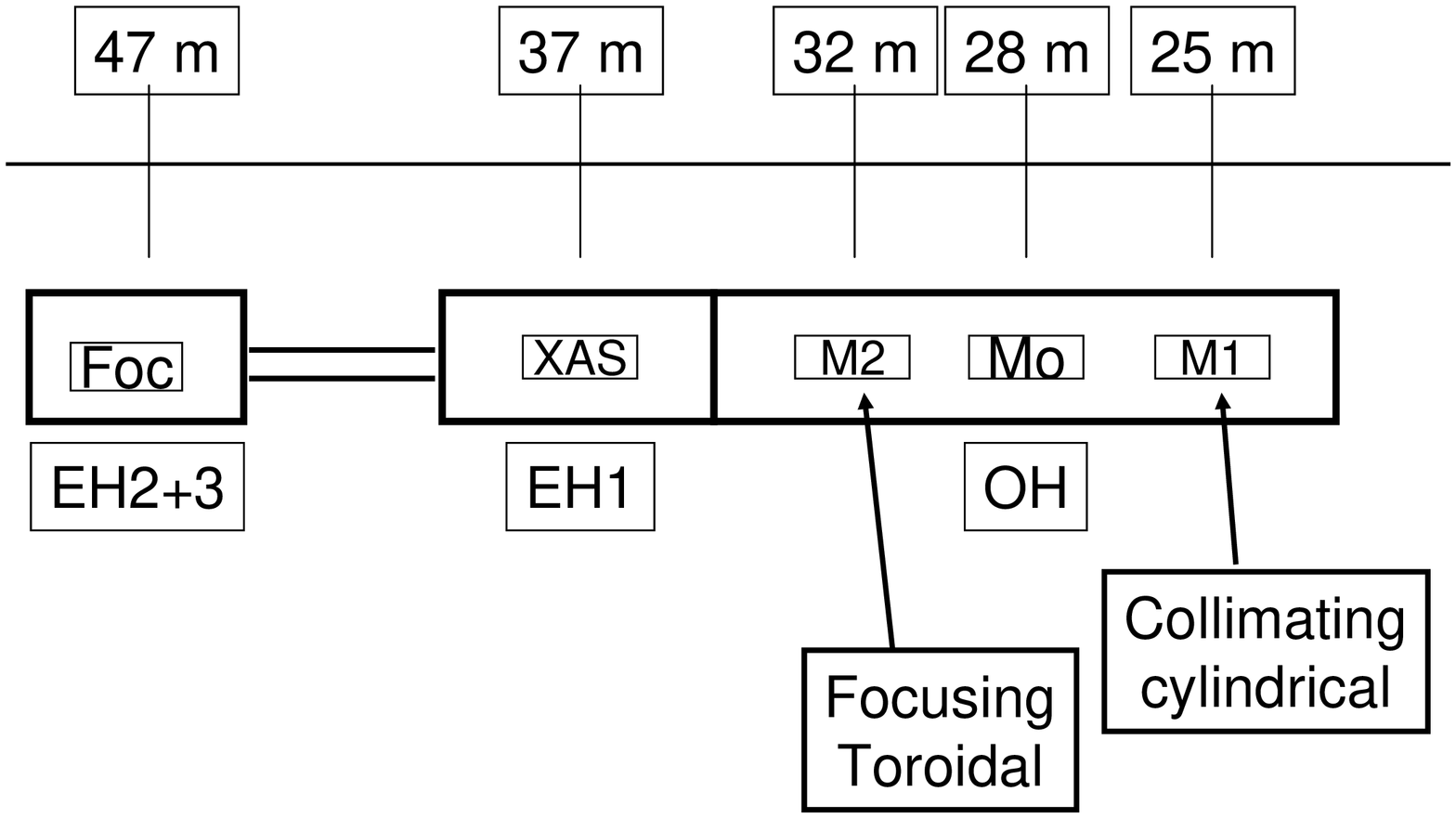}
\caption [Sketch of the beamline]{Sketch of the proposed beamline.
In the upper part of the picture the distance of the various
elements from the source is indicated.} \label{fig:bl-sketch}
\end{center}
\end{figure}

The layout of the beamline is shown in Fig.\ref{fig:bl-sketch}: and will consist in 3 lead cabins: an optical hutch (OH) and two experimental hutches (EH1 and EH2) The Optical Hutch will contain the optical elements that will focus the beam in the middle of the second experimental hutch EH2. Here, there will be the main experimental setup consisting in two stations, one with a vacuum chamber (containing a manipulator for standard measurements and one for ReflEXAFS experiments) and the other will be an open station permitting the mounting of bulky sample environments and detection systems.The two setups will be mounted on a long marble bench and will have the possibility of translate along the beam path to match the the sample location with the focal position. The preceding hutch (Experimental Hutch 1, EH1) will be used for experiments  with a non focused beam like EXAFS in transmission mode and low noise conditions. \\
The X-ray optics scheme will consists in a collimating first mirror a double crystal monochromator and a focusing second mirror. The beamline will use mirrors in the energy range 5-40 keV. Above 40 keV the monochromator will be the only optical element. \\

\subsection{X-ray source}
The beamline will take the beam from the high field part (0.85 T) of the BM8 bending magnet of ESRF. In this point the X-ray source has dimensions $75 \mu m$ hor and $32 \mu m$ vert. Full Width at Half Maximum. The source is located at 25 m from the first optical element. In the following we will consider the machine running at 6 GeV with a stored current of 200 mA. The realization of the new lattice of ESRF \cite{esrf-white-paper} will lead to a smaller source ($13 \times 4 \mu m$) and a movement of it 2.5 m farther. The design is such that it will be possible to cope with these changes and that it will profit at maximum of the foreseen new features. 

\subsection{Mirrors}
The problem of beam focusing is a major issue for the present project. To meet the requirements stated in the scientific case it is necessary to have at the same time high brilliance, flux and beam homogeneity that are difficult to meet at the same time. On other beamlines different optical schemes have been adopted to focus the beam. In some cases sagittal focusing has been chosen (FAME at ESRF \cite{FAME}, SAMBA \cite{SAMBA} and DIFFABS \cite{DIFFABS} at SOLEIL) . Sagittal focusing, although providing exceptionally intense beams, is not suited for our goals as it introduces severe beam inhomogeneities in the horizontal plane. Good results have been obtained from non-focusing designs (BM29 at ESRF \cite{bm29}, XAFS beamline at ELETTRA \cite{xafs-elettra}); but this technical solution would restrict the application field of the beamline to concentrated samples that is too limiting for the scientific case presented here. In more recent projects a great interest for toroidal mirrors has emerged driven by the achromaticity of this element and the strong improvement of the quality of these devices in the latest years. This is the case of protein crystallography on bending magnet beamlines at APS \cite{04-jsr-mcdowell}, B18 at DIAMOND \cite{B18}, the ROBL beamline at ESRF \cite{ROBL}, the CLAESS beamline at ALBA \cite{CLAESS}. This choice will be chosen for the present project as it ensures at the same time a high intensity and homogeneous beam with a reduced size. \\
The first mirror has a cylindrical shape and is used for collimating the beam as it permits to lower the vertical divergence of the beam from $\approx 40 \mu rad$ (determined by the input slits) to the limit of the mirror slope error (currently achievable: a few $\mu rads$). As the scattering vector of the monochromator is vertical, this device permits to achieve an instrumental energy resolution well below (less than 25\%) the core-hole width of the K edges in the energy range of interest. This will realize an optimized configuration for the collection of XANES data at high energy resolution. This mirror will have 3 stripes: Pt, Pd and Si in order to cover the whole energy range harmonic free. Si will cover the region 5-15 keV whereas Pt will cover 15-40 keV so ensuring an harmonic-free operation in this total energy range. Pd will cover the region around 15 keV in order to avoid the cutoff of the Si coating and the absorption lines from the Pt coating. \\
The second mirror will have a toroidal shape and will focalize the beam in a 2:1 condition on the horizontal plane. It will consist in a cylindrical channel cut in a bendable silicon substrate. This choice has been shown to minimize the impact of comatic aberrations in the focal spot \cite{04-jsr-mcdowell} and will keep at a moderate value the overall length of the beamline. The limited horizontal acceptance of the device (limited to 1mrad in this case) is largely compensated by the high quality of the beam achievable in this way. Considered the space available it will be possible to have 2 toroidal channels on the same substrate. The part between the channels section of the second mirror will be left flat to make available to users a non focused beam. The technical solution consisting in 2 channels realized in the same substrate is not particularly risky and has been successfully used in other projects (\cite{CLAESS}, \cite{B18}, \cite{ROBL}). The two channels will be coated one with Pt and the other with Pd, the flat region between the channels will be used for applications needing unfocused beam. \\
Both mirrors will work at 2 mrad incidence angle.  For focusing above 40 keV a solution using graded multilayers is at present under study.  Table \ref{tab:mirrors} collects the principal parameters of these optical elements. \\
A pair of Pt-coated plane mirrors (already present on the beamline) working at 10mrad will be used to reject harmonics in the lowest part of the energy range. 

\begin{table*}
\caption{Geometric parameters of the mirrors.}
\label{tab:mirrors}
\begin{tabular}{|c|c|c|c|}
  \hline
  Mirror & Shape & Major Radius & Minor Radius \\
  1 & Cylindrical & 24.8 Km & - \\
  2 & Toroidal & 15.8 Km & 4.2 cm \\
  \hline
\end{tabular}
\end{table*}
\subsection{Monochromator}
The monochromator will be a fixed-exit device with a single rotation axis and will use flat crystals. The first crystals will be cooled with water. Different pairs of crystals will be mounted to access the whole energy range: Si(311) will cover the interval 5-30 keV,  whereas Si(511) will operate in the range 10-80 keV. A further Si(111) crystal will be used to obtain high photon flux in the region 5-15 keV. The crystals will be permanently mounted inside the monochromator and the change will be realized by simple horizontal translation of the instrument. This solution will ensure a high quality beam in terms of temporal stability and beam spatial homogeneity thanks to the use of flat reflectors. Depending on the details of the design (in particular, the max and min Bragg  angle achievable) the operation energy range of the various crystals could be modified. \\
\subsection{Other elements}
Before the first mirror and in analogy with the design of the GILDA beamline a set of 9 attenuators will be mounted. This will permit at all energies the reduction of the thermal load on the subsequent optical elements by absorbing the unnecessary low energy part of the spectrum. In particular with this element we aim in reducing at maximum the thermal bump expected on the first mirror or on the first crystal when operating above 40 keV and that otherwise could severely degrade the energy resolution and the size of the focal spot.  \\
A set of cooled primary slits will be mounted after the attenuators and further slits will be placed before and after the monochromator. Before and after each optical element a beam monitor consisting in a fluorescent screen and a linear detector (photodiode or W wire) will be installed for beam characterization.
\subsection{Ray Tracing}
\begin{figure}
\begin{center}
\includegraphics[width=10cm, angle = -90] {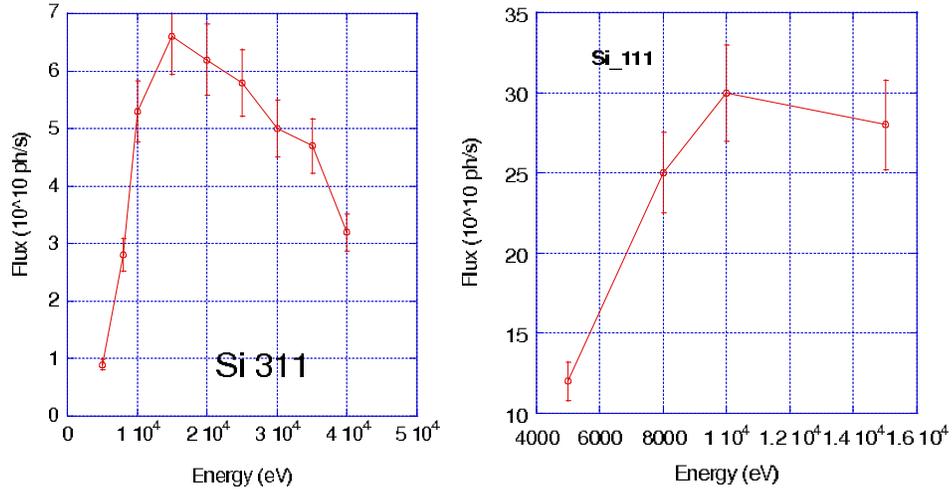}
\caption [Sketch of the beamline]{Flux foreseen on the focal spot in the case of two possible crystal planes for the monochromator. The simulation condition are those mentioned in Tab.~\ref{tab:beamline}.}
\label{fig:flux}
\end{center}
\end{figure}

\begin{figure}
\begin{center}
\includegraphics[width=10cm] {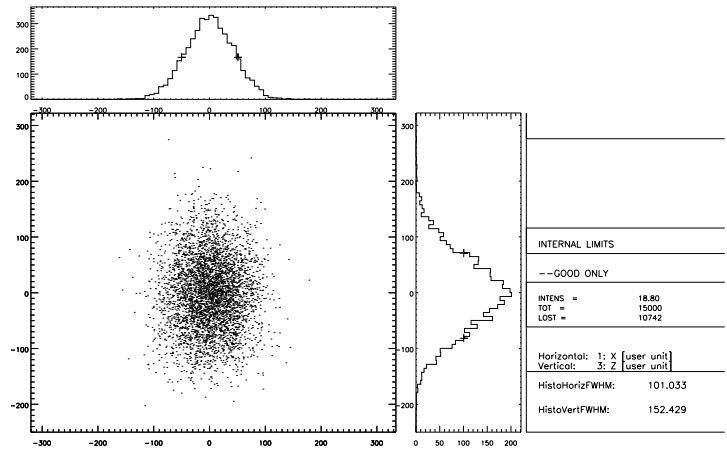}
\caption [Focal spot]{Geometry of the focal spot}
\label{fig:spot}
\end{center}
\end{figure}

A detailed Ray-Tracing calculation of the parameters of the beamline has been carried out using the SHADOW code \cite{86-nim-lai}. The beamline is shown to provide a high photon flux in a wide energy range and a small beam size in compliance with the experimental requirements cited in the previous sections (See Fig.~\ref{fig:flux}).
Table \ref{tab:beamline} collects all the main parameters derived from the simulation:
\begin{table*}
\caption{General parameters of the beamline as calculated by ray-tracing. Real values of slope error, about 2$\mu$rad of the mirrors have been accounted for in the calculation. Machine parameters were a current of 200mA and a beam acceptance was 1mrad H * 0.043 mrad V.}
\label{tab:beamline}
\begin{tabular}{|c|c|}
\hline
Parameter & Value \\
\hline
Operating Energy range & 5.4-70 keV \\
Flux w Si(311) & 1-7 $10^{10}$ ph/s  \\
Flux w Si(111) & 1-3 $10^{11}$ ph/s  \\
Energy resolution w Si(311) & $4*10^{-5}$ - $1*10{-4}$ \\
Energy resolution w Si(111) & $1*10^{-4}$ - $1.5*10{-4}$ \\
Beam size H *V FWHM & 103 * 143 $\mu$m \\
Beam Divergence H * V & 0.050 * 0.004 deg \\
\hline
\end{tabular}
\end{table*}
An example of the geometry of the focal spot is presented in Fig.~\ref{fig:spot}.
\subsection{Endstations}
The beamline will have 2 experimental stations: the main one, using the focused beam, will be  placed in the EH2 . The main station in EH2 will be a long marble bench carrying 2 experimental stations and supports for the ion chambers, one measuring the incoming beam ($I_0$) and the others reading beam transmitted through the sample ($I_1$) and the reference ($I_R$). The first experimental stations will carry a vacuum chamber containing two manipulators: one for standard XAS experiments in fluorescence mode and the other for ReflEXAFS experiments. Sideways, a Be window will permit the X-ray arrrival to the Fluorescence detector. The second station will be a table with basic translation movements. This will be used to mount bulky experimental setups and multitechnique detection systems.\\
A new multielement Ge detector will be a keypoint of the beamline. It will be a 25 elements device with an almost doubled sensitive surface respect to the previous detector and will increase dramatically the capability of investigating diluted samples. 
In EH1 there will be an apparatus for the collection of conventional XAS spectra in transmission mode. It will consist in a bench hosting two experimental chambers and ion chambers for the detection of the beam before and after the sample.Used in conjunction to a non-focusing configuration of the beamline it will realize a station for high quality XAS data collection. \\
The ancillary equipment will include a liquid helium/nitrogen cryostat, a cell for chemical reactions gas-solid and a cell for measurements in liquid phase. For the control of the optical elements and instruments in OH and EHs the standard ESRF hardware and software will be used with a consequent ease of operation for users.
\subsection{Infrastructures}
	In order to accomodate the new endstations and the new beam path the existing infrastructures (hutches, lead walls, electrical cabling) need to be revised. In particular it will necessary to eliminate the lead wall between the present EH2 and EH3 (in order to create a new large EH2) and to adapt the height of the beam pipes and pass-through lead shieldings to the new value. The electrical cabling will be consequently revised as well as the fluid pipes in the old EX2 and 3 and the PSS system.
\subsection{Day 1 performance}
In order to satisfy the needs of the user community and to be competitive respect to other projects
active in the other synchrotrons we aim in realizing a beamline that will have the following
 \textit{day 1 performance} (See Tab.~ \ref{Tab:d1p}). \
\begin{table*}
\caption{Target day 1 performance of the beamline}
\label{Tab:d1p}
\begin{tabular}{|c|c|}
  \hline
 Energy range \ Focused & 5-40 keV \\
 Energy range \ Unfocused & 5-80 keV \\
 Flux \ Focused& $\geq 10^{10}$ ph/s \\
 Flux \ Unfocused& $\geq 10^{8}$ ph/s \\
 Beam size & $\leq 200*200 \mu m^2 $ FWHM \\
Noise on transmission XAS & $ < 10^{-4}$  \\
  \hline
\end{tabular}
\end{table*}
\newpage
\section{Long Term Perspectives}
In this new layout GILDA will become a beamline with a high quality beam thanks to the optical scheme based on mirror focusing. This scheme reveals to be particularly well adapted to the configuration that will be realized at ESRF after the completion of the Phase II of the Upgrade \cite{esrf-up-ph2}. In particular the availability of mini insertion device (3 poles wigglers) sources in place of the previous bending magnets will increase of at least a factor 3 the flux on the sample maintaining a narrow horizontal emission angle and exploiting at their best the focusing characteristics of the toroidal mirrors. With this new source the CRG beamlines at ESRF will become considerably attractive and GILDA in her new configuration will constitute a unique occasion of improved and new experiments for the italian and international users communities. 
\clearpage


\end{document}